\documentclass[%
 reprint,
superscriptaddress,
 amsmath,amssymb,
 aps,
pra,
floatfix,
]{revtex4-1}
\usepackage{graphicx}
\usepackage{dcolumn}
\usepackage{bm}
\usepackage{booktabs}
 
 \usepackage{array}
 \usepackage[T1]{fontenc}

\begin{document}

\hyphenpenalty=2000
\preprint{APS/123-QED}

\title{High-harmonic generation enhanced by laser-induced electron recollision}
\author{Yang Li}
\affiliation{\text{Department of Nuclear Engineering and Management, Graduate School of Engineering,}
The University of Tokyo, 7-3-1 Hongo, Bunkyo-ku, Tokyo 113-8656, Japan}

\author{Takeshi Sato}
\email{sato@atto.t.u-tokyo.ac.jp}
\affiliation{\text{Department of Nuclear Engineering and Management, Graduate School of Engineering,}
The University of Tokyo, 7-3-1 Hongo, Bunkyo-ku, Tokyo 113-8656, Japan}
\affiliation{\text{Photon Science Center, Graduate School of Engineering,}
The University of Tokyo, 7-3-1 Hongo, Bunkyo-ku, Tokyo 113-8656, Japan}
\affiliation{\text{Research Institute for Photon Science and Laser Technology,}
The University of Tokyo, 7-3-1 Hongo, Bunkyo-ku, Tokyo 113-8656, Japan}

\author{Kenichi L. Ishikawa}
\email{ishiken@n.t.u-tokyo.ac.jp}
\affiliation{\text{Department of Nuclear Engineering and Management, Graduate School of Engineering,}
The University of Tokyo, 7-3-1 Hongo, Bunkyo-ku, Tokyo 113-8656, Japan}
\affiliation{\text{Photon Science Center, Graduate School of Engineering,}
The University of Tokyo, 7-3-1 Hongo, Bunkyo-ku, Tokyo 113-8656, Japan}
\affiliation{\text{Research Institute for Photon Science and Laser Technology,}
The University of Tokyo, 7-3-1 Hongo, Bunkyo-ku, Tokyo 113-8656, Japan}


\begin{abstract}
We theoretically investigate correlated electron dynamics in high-harmonic generation (HHG), using all-electron \emph{ab initio} simulations for three-dimensional real alkali-metal atoms. The resulting harmonic spectra exhibit a plateau extended beyond the usual cutoff and a prominent resonance peak above the plateau. These remarkable features arise from the cation response dramatically enhanced by laser-induced electron recollision, which is a key process of attosecond science. This demonstrates that high-harmonic spectroscopy provides new possibilities to explore dynamical electron correlation in strong laser pulses. 
\end{abstract}

\maketitle
\section{INTRODUCTION} \label{INTRODUCTION}
Atoms, molecules, and solids subject to intense laser pulses exhibit highly nonlinear phenomena known as strong-field phenomena. High-harmonic generation (HHG), in particular, serves as a highly successful means to generate coherent attosecond pulses in the extreme ultraviolet and soft x-ray regions \cite{PMPaul,MHentschel,EJTakahashi,TGaumnitz}. It has given birth to attosecond science \cite{PAgostini,PBCorkum}, as an essential tool to initiate and probe ultrafast electron dynamics \cite{JItatani,PSalieres,OSmirnova,FCalegari,PMKraus2}.
High-harmonic spectroscopy crucially relies on how the electronic structure and dynamics are embedded in HHG spectra.
For example, harmonic spectra reflect the Cooper minimum in $\mathrm{Ar}$ \cite{HJWorner,JHiguet}, autoionizing resonance in transition-metal plasma plumes \cite{SHaessler,SKlumpp,MAFareed}, and the giant resonance in $\mathrm{Xe}$ \cite{ADShiner}. All of these are basically understandable as features of single-photon ionization, which is the inverse process of recombination, the last step in the three-step model \cite{PBCorkum2,KCKulander} of HHG.
At the same time, the last two examples suggest the importance of multielectron effects in HHG \cite{TSato2}.
Recently, we \cite{ITikhomirov} and subsequently Abanador {\it et al.}~\cite{PMAbanador} have found that excitation of the remaining ionic core by laser-induced electron recollision, a key process in attosecond science with strong electron correlation, can in principle induce enhanced resonant peak and second-plateau formation in HHG spectra.
However, since these studies have used one-dimensional model systems, it is not clear if such a drastic effect takes place to an experimentally observable extent in real systems.

In this paper, we revisit this problem using state-of-the-art, fully {\it three-dimensional} all-electron first-principles simulations and predict that the above-mentioned new mechanism in HHG is real.
We numerically simulate HHG from alkali-metal atoms ($\mathrm{Li}$ and $\mathrm{Na}$) in intense few-cycle mid-infrared (MIR) laser pulses ($\lambda=1200$ nm), utilizing the multiconfiguration time-dependent Hartree-Fock (MCTDHF) method \cite{KLIshikawa,JZanghellini,TKato,JCaillat,GJordan,RSawada} in its full dimensionality. The MCTDHF is based on the configuration-interaction expansion of the total wave function using time-dependent variationally optimized orbitals, and thus can provide first-principle framework of tracking dynamical electron correlation. We find the harmonic spectra of both $\mathrm{Li}$ and $\mathrm{Na}$ indeed exhibit double-plateau structures. While the cutoff position of the first plateau is consistent with the prediction of the semiclassical three-step model \cite{PBCorkum2,KCKulander} for the neutral species, the second plateau is recognized to be from the contribution of the cations ($\mathrm{Li^+}$ and $\mathrm{Na^+}$). In addition, a prominent peak shows up in the spectrum of $\mathrm{Li}$, whose position coincides with the resonant excitation spectrum of $\mathrm{Li^+}$. Our analyses reveal that these both originate from laser-induced electron recollision that dramatically enhances harmonic response of the cations, hence, a clear manifestation of dynamical electron correlation.

This paper is organized as follows. In Sec.~\ref{MCTDHF}, we describe the MCTDHF method and the computational details. In Sec.~\ref{DISCUSSION}, numerical results for the ionization dynamics and high-harmonic response of alkali metal atoms are presented. Conclusions are given in Sec.~\ref{CONCLUSION}.
Atomic units are used throughout unless otherwise stated.

\section{METHOD} \label{MCTDHF}
We consider a fully correlated, $N$-electron atom (or ion) with atomic number $Z$ in a laser field $\mathbf{E}(t)$ linearly polarized along $z$ direction. In the velocity gauge, the total Hamiltonian reads,
\begin{eqnarray}
\hat{H}(t)&=&\sum_{j=1}^N\left(-\frac{\nabla_j^2}{2}-\frac{Z}{r_j}-i\mathbf{A}(t)\cdot\mathbf{\nabla}_j\right)\nonumber\\ 
& &+\sum_{j=1}^{N-1}\sum_{k=j+1}^{N}\frac{1}{|\mathbf{r}_j-\mathbf{r}_k|}, 
\label{eqtdse}
\end{eqnarray}
where $\mathbf{A}(t)=-\int_{-\infty}^{t}\mathbf{E}(t')dt'$ denotes the vector potential of the laser pulse and $\mathbf{r}_j$ ($j=1,\cdots,N$) the spatial coordinate of the $j$th electron. Within the MCTDHF treatment, the normalized multiconfiguration $N$-electron wave function is expressed as a linear combination of Slater determinants $\Phi_I(t)$,
\begin{eqnarray}
\Psi(t)=\sum_{I}\Phi_I(t)C_I(t),
\label{eqmctdhf}
\end{eqnarray}
where $C_I(t)$ is the configuration-interaction (CI) coefficients. The Slater determinants $\Phi_I(t)$ are constituted by an orthonormal set of $2M$ spin orbitals $\left\{\psi_i^\sigma(t)\right\}$ ($i=1,\cdots,M$; $\sigma=\alpha,\beta$), with $\psi_i^\sigma=\phi_i\otimes|\sigma\rangle$, where $\left\{\phi\right\}$ and  $\left\{\sigma\right\}$ are the single-electron spatial orbital functions and spin eigenstates, respectively. The summation in Eq.~(\ref{eqmctdhf}) runs over all possible permutations among $2M$ spin orbitals. Both the CI-coefficients and the Slater determinants are varied in time. The equations of motion for the CI-coefficients and orbital functions are derived by utilizing the time-dependent variational principle \cite{JZanghellini,TKato,JCaillat,GJordan,RSawada,TSato,TSato2}. 
Our numerical implementation have been detailed in Refs.~\cite{TSato2,YOrimo}.

Specifically, we consider $\mathrm{Li}$ ($N=Z=3$) and $\mathrm{Na}$ ($N=Z=11$) atoms under the irradiation of a few-cycle MIR laser pulse with a wavelength of $\lambda=1200$ nm, a peak intensity of $2\times10^{14}$ $\mathrm{W/cm^2}$ and the foot-to-foot pulse duration of 8$T$ with $T$ being the optical cycle. The electric field $\mathbf{E}(t)$ is defined as $\mathbf{E}(t)=E_0\sin^2(\pi t/8T)\sin(\omega t) \mathbf{e}_z$. We use up to 20 spin orbitals ($M=10$) for $\mathrm{Li}$ and 30 ($M=15$) for $\mathrm{Na}$, with which the harmonic spectra are converged with respect to $M$. The spherical finite-element discrete variable representation (FEDVR) \cite{TNRescigno} is adopted for the description of spatial orbital functions. We use 40 radial finite elements with 23-point DVR per FE. An efficient absorbing boundary called infinite-range exterior complex scaling (irECS) \cite{AScrinzi,YOrimo} is placed at $R=148$ a.u., which is about three times larger than the quiver radius of $\sim 52$ a.u. and thus large enough to accommodate the electron excursion during the laser pulse.  For real-time propagation, each orbital is expanded with 71 spherical harmonics and the time step is $\Delta t=0.005$ a.u. We have confirmed that all the results are converged with respect to the radial, angular and temporal resolutions. The initial ground states are obtained by imaginary-time propagation. The calculated total ground state energies are -7.4764 a.u. for $\mathrm{Li}$ and -162.0792 a.u. for $\mathrm{Na}$, in good agreement with the NIST values of -7.4780 a.u. and -162.4322 a.u. \cite{AKramida}. HHG spectra are computed as the squared modulus of the Fourier transformed electric dipole acceleration. In order to reduce the background level of the HHG spectra while maintaining any possible excitation signals, the propagation is continued for four additional optical cycles after the end of the pulse and the dipole acceleration is multiplied by a Dolph-Chebyshev window function before the Fourier transform.

\section{RESULTS AND DISCUSSIONS} \label{DISCUSSION}
\begin{figure}[tb]
\centerline{\includegraphics[width=8.5cm,angle=0,clip]{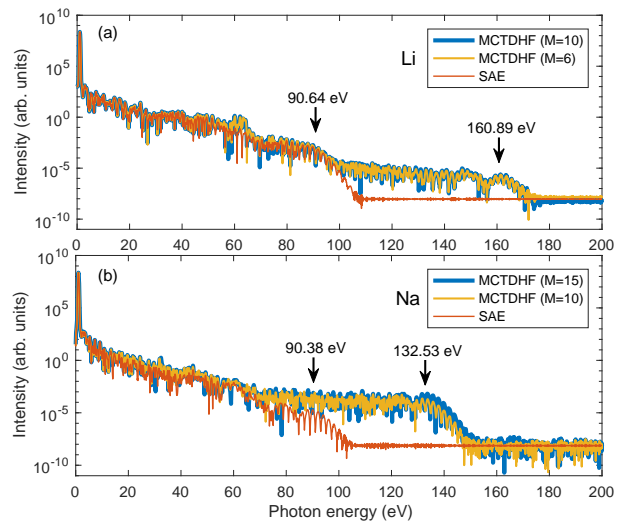}}
\caption{High-harmonic spectra from (a) $\mathrm{Li}$ and (b) $\mathrm{Na}$, calculated using the MCTDHF method with different number $M$ of orbitals and using the SAE approximation. See text for the laser parameters. The arrows indicate the positions of cutoff energies listed in Table \ref{table1}. } 
\label{Figure1}
\end{figure}

Figure~\ref{Figure1}(a) and~\ref{Figure1}(b) show the harmonic spectra of $\mathrm{Li}$ and $\mathrm{Na}$, respectively, with different number of spatial orbitals. One can clearly see that the convergence of the HHG spectra is achieved with increasing the number of orbitals. While for $\mathrm{Li}$ $M=6$ suffices to essentially reproduce the fully converged spectrum ($M=10$), up to $M=10$ is needed for $\mathrm{Na}$ to reach overall good agreement with the most accurate result ($M=15$). To the best of our knowledge, the fully converged results presented here are the first full-dimensional \emph{ab initio} simulation of HHG in the MIR ($\sim1200$ nm) high-intensity ($\sim10^{14}$ $\mathrm{W/cm^2}$) regime. The multielectron spectra exhibit two remarkable features: (i) a double-plateau structure with a second plateau extended far beyond the cutoff of the main plateau for both $\mathrm{Li}$ and $\mathrm{Na}$ and (ii) a noticeable peak around 62 eV for $\mathrm{Li}$. For comparison, HHG spectra within the single-active-electron (SAE) approximation are also presented in Fig.~\ref{Figure1}. The SAE spectra are calculated with the time-dependent complete-active-space self-consistent-field (TD-CASSCF) method \cite{TSato,TSato2}, by treating only one valence electron as \textit{active} and all the other (inner-shell) orbitals as \textit{frozen core}. Whereas the SAE results show overall agreement with the MCTDHF spectra in the first-plateau region, they fail to reproduce the peak ($\sim 62\,{\rm eV}$) of $\mathrm{Li}$ and the second plateau. This observation unambiguously indicates that the inner-shell electrons are involved and make critical contributions in the HHG process.

\begin{table}[tb]
	\caption{Ionization potential $I_p$ evaluated through Koopmans' theorem, which agrees with the experimental value within the displayed digits, and the HHG cutoff energy $E_c$ for different species.}
	\begin{ruledtabular}
	\renewcommand{\arraystretch}{1.5}
		\begin{tabular}{lccccr}
			 & $\mathrm{Li}$ & $\mathrm{Li^+}$ & $\mathrm{Na}$ & $\mathrm{Na^+}$\\
			\hline
			$I_p$ (eV)&  5.39  &  75.64  &  5.14  &  47.29\\
			$E_c$ (eV)&  90.64 &  160.89 &  90.38 &  132.53\\
		\end{tabular}
	\end{ruledtabular}
	\label{table1}%
\end{table} 

Let us analyze the cutoff positions of the MCTDHF spectra. In Table \ref{table1}, we list the ionization potential $I_p$ and the cutoff energy $E_c=I_p+3.17U_p$ predicted by the three-step model \cite{PBCorkum2,KCKulander}, with $U_p$ being the ponderomotive energy, for both neutral atoms and cations.  The positions of $E_c$ are indicated in Fig.~\ref{Figure1} by arrows. One can see that the cutoff energies of the second HHG plateau coincide with those for $\mathrm{Li^+}$ and $\mathrm{Na^+}$, which suggests the second plateau are from the contribution of the cations. To validate this speculation, we introduce the spatial-domain-based charge-state-resolved dipole acceleration, to effectively divide the total harmonic signal into the contributions of different charge states. It is defined, for the charge state $n+$, as \cite{TSato},
\begin{eqnarray}
a_n(t)&\equiv&\binom Nn\int_>dx_1\cdots\int_>dx_n\int_<dx_{n+1}\cdots\int_<dx_N \nonumber\\
& &\times \Psi^\ast(x_1,\cdots,x_N,t)\ddot{z}\Psi(x_1,\cdots,x_N,t),
\label{csrda}
\end{eqnarray}
where $\int_>$ and $\int_<$ denote integrations over the spatial-spin coordinate $x=\left\{\mathbf{r},\sigma\right\}$ with the spatial part restricted to the regions $\left|\mathbf{r}\right|>R_0$ and $\left|\mathbf{r}\right|<R_0$, respectively, with $R_0=20$ a.u. characterizing the boundary of electron ionization. The explicit form of acceleration operator $\ddot{z}$ is described in Ref. \cite{TSato2}. 

\begin{figure}[tb]
\centerline{\includegraphics[width=8.5cm,angle=0,clip]{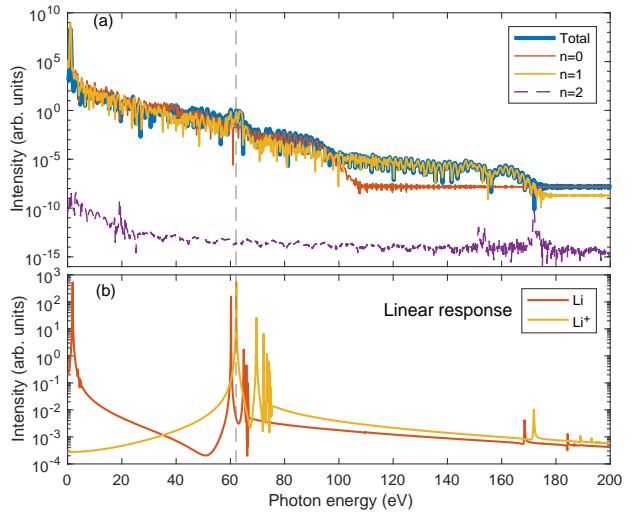}}
\caption{(a) Charge-state-resolved harmonic spectra calculated by Fourier transform of Eq.~(\ref{csrda}) with $n=0, 1, 2$. The total spectrum, same as the yellow curve in Fig.~\ref{Figure1}(a), is also shown for comparison. (b) Excitation spectra of $\mathrm{Li}$ and $\mathrm{Li^+}$, calculated through excitation by a quasi-delta pulse. The strongest line, corresponding to the $1s^2\rightarrow 1s2p$ excitation of $\mathrm{Li^+}$, is located at 62.20 eV. This value is in excellent agreement with the NIST data of 62.22 eV \cite{AKramida}. The grey vertical dashed line guides the position of the peak.}
\label{Figure2}
\end{figure}

Figure~\ref{Figure2}(a) shows the change-state-resolved harmonic spectra for $\mathrm{Li}$ with $M=6$. The first plateau in the total spectrum, except for the peak around 62 eV, is mainly formed by the response of $n=0$, which is similar to the SAE spectrum shown in Fig.~\ref{Figure1}(a). As deduced above, the second plateau originates from the contribution of $n=1$. The harmonic signal of $n=2$ is orders of magnitude lower and hence makes no contribution to the total spectrum. Another noticeable feature is that the prominent peak at around $62$ eV in the first plateau region is well reproduced by the cation response, which implies possible excitation resonance of $\mathrm{Li^+}$. We thus calculate the linear response of $\mathrm{Li}$ and $\mathrm{Li^+}$ to a quasi-delta pulse with an intensity of  $10^{10}$ $\mathrm{W/cm^2}$. Note that unlike Fig.~\ref{Figure2}(a), this calculation starts from the ground state of $\mathrm{Li}$ and $\mathrm{Li^+}$, respectively. The excitation spectra are obtained by Fourier transform of the dipole, as presented in Fig.~\ref{Figure2}(b). Around 62 eV, there are two dominant spectral lines in the excitation spectra. One is associated with $1s^22s\rightarrow1s2p^2$ double excitation of $\mathrm{Li}$ at 60.24 eV (experimental value from NIST: 60.81 eV \cite{AKramida}), and the other corresponds to $1s^2\rightarrow 1s2p$ excitation of $\mathrm{Li^+}$ at 62.20 eV (experimental value from NIST: 62.22 eV \cite{AKramida}). We attribute the peak in the harmonic spectrum to the latter, since no resonance peak is found in the harmonic response from $n=0$, as shown by the red curve in Fig.~\ref{Figure2}(a).

\begin{figure}[tb]
\centerline{\includegraphics[width=8.5cm,angle=0,clip]{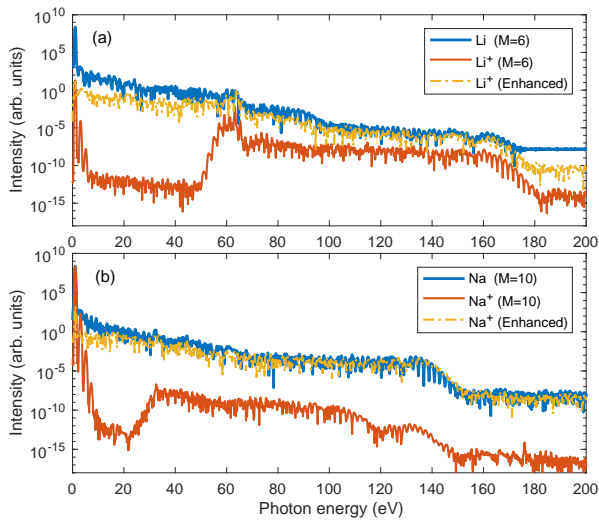}}
\caption{(a) High-harmonic spectrum from $\mathrm{Li^+}$ (red solid curve) using the MCTDHF method with 6 orbitals, compared with HHG spectrum from $\mathrm{Li}$ [blue solid curve, the same as in Fig.~\ref{Figure1}(a)] with the same number of orbitals. The dashed-dotted yellow curve is the spectrum calculated starting from $\mathrm{Li^+}$ but including the dynamical electron correlations (see text for details). (b) Same as (a) but for $\mathrm{Na}$ and $\mathrm{Na^+}$.}
\label{Figure3}
\end{figure}

Up to now, we have shown that both the unusual features in the MCTDHF spectra, the resonance peak and the second plateau, require consideration of the core electrons. A simple explanation might be single ionization of the neutral followed by a sequential HHG process in the cation \cite{YAkiyama}. The second cutoff exceeds the usual atomic cutoff due to the high ionization potential of the cation and the resonance peak might be caused by the laser excitation of the cation.  To examine the plausibility of this scenario, we calculate the HHG spectra starting from $\mathrm{Li^+}$ and $\mathrm{Na^+}$ by the MCTDHF method using the same laser pulse as in Fig.~\ref{Figure1}.  The comparison between neutral (blue, the same as shown in Fig.~\ref{Figure1}) and ionic (red) harmonic spectra for both $\mathrm{Li}$ and $\mathrm{Na}$ are shown in Fig.~\ref{Figure3}. Most noticeably, whereas the cutoff positions of the ionic spectra match the second cutoffs of the neutral, the strength of the plateau signals is several orders of magnitude weaker than the neutral spectra. Moreover, although the $\mathrm{Li^+}$ response also exhibits a resonance peak at around 62 eV, the intensity is too low to explain the peak in the spectrum of $\mathrm{Li}$. These observations indicate that the sequential mechanism, without consideration of the dynamical correlations between energetic recolliding electron and core electrons, fails to explain the intensity of the second plateau and the resonance peak.

In order to explore the underlying mechanism inducing the drastic enhancement of the second plateau and the resonance peak, we develop an intuitive approach that incorporates the electron-electron correlation during the interaction with the laser pulse. We first record the instantaneous probability density $\rho(\mathbf{r},t)=|\Psi(\mathbf{r},t)|^2$ of the active electron in the SAE calculations of the neutral. The motion of the active electron forms an oscillating dipole, interacting with the inner electrons (or those in the parent ion) through Coulomb force, which is described by a potential,
\begin{eqnarray}
\hat{V}_{\mathrm{cor}}=\sum_{j=1}^K\int d\mathbf{r'}\frac{\rho(\mathbf{r'},t)}{|\mathbf{r}_j-\mathbf{r'}|},
\label{cor}
\end{eqnarray}
where $K$ is the number of the inner electrons ($K=2$ for $\mathrm{Li}$ and 10 for $\mathrm{Na}$). Note that Eq.~(\ref{cor}) explicitly characterizes the dynamical electron correlation between the ejected electron and the other electrons. The only approximation is that the exchange effect is neglected. We then add this term to the Hamiltonians Eq.~(\ref{eqtdse}) for $\mathrm{Li^+}$ and $\mathrm{Na^+}$, and calculate the cation harmonic responses in the laser field. The resulting spectra are shown by dashed-dotted yellow curves in Fig.~\ref{Figure3}. Compared to the response of the bare cation (red), the intensity of the spectra is substantially enhanced, to the level of solid blue curves. Moreover, the resonance peak in the spectrum of $\mathrm{Li}$ is also recovered. This result unambiguously demonstrates that the resonance peak and the second plateau are augmented to the observable level as a result of dynamical electron correlation. In the sense of the three-step model, for $\mathrm{Li}$, the recolliding first electron populates $\mathrm{Li^+}$ to the first excited state $1s2p$, which leads to extreme ultraviolet photon emission, forming the peak at 62.2 eV.  At the same time, electron is ejected from the excited state and recombines to the ground state to generate high harmonics. Due to the low ionization potential, i.e., high tunneling rate of the excited state, the intensity of the harmonics in the second plateau is greatly enhanced. For $\mathrm{Na}$, the oscillating dipole also virtually excites $\mathrm{Na^+}$ and facilitates following harmonic generation in the second plateau. While this mechanism is similar to double-plateau formation from a coherent superposition state \cite{JBWatson,ASanpera} and enhancement by an assisting short-wavelength pulse \cite{KLIshikawa2,KLIshikawa3,EJTakahashi,KLIshikawa4}, the enhancement here is due to dynamical electron correlation, i.e., the direct Coulomb force from the recolliding electron.  

\begin{figure}[tb]
\centerline{\includegraphics[width=9cm,angle=0,clip]{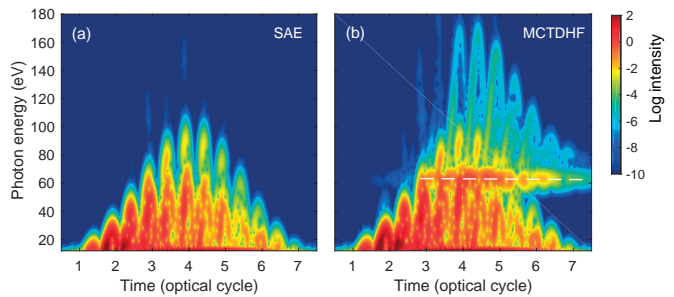}}
\caption{\label{fig:Gabor} Time-frequency Gabor analysis of the (a) SAE and (b) MCTDHF dipole accelerations for $\mathrm{Li}$, corresponding to the harmonic spectra shown in Fig.~\ref{Figure1}(a). The size of the Gaussian window function in the Gabor analysis is 0.1 optical cycle. The white horizontal dashed line indicates the strong spectral emission around 62.2 eV.}
\end{figure}

Finally, we perform time-frequency analysis of HHG for $\mathrm{Li}$ by Gabor transform of the dipole acceleration (Fig.~\ref{fig:Gabor}). In the SAE case [Fig.~\ref{fig:Gabor}(a)], high-harmonics are mostly generated in the rising edge of the laser pulse, which is reasonable in view of the high intensity of the laser pulse. In Fig.~\ref{fig:Gabor}(b), the signal of photon emission contributing to the second plateau and the resonance peak occurs starting from $t\approx3T$, which is delayed compared to that of the first plateau. This clearly reveals that HHG from the inner electrons is initiated by dynamical electron correlation when the energy of the returning electron exceeds the cation excitation energy.

\section{CONCLUSIONS} \label{CONCLUSION}
In conclusion, we have performed full-dimensional, all-electron \emph{ab initio} investigations on high-harmonic generation of alkali-metal atoms  ($\mathrm{Li}$ and $\mathrm{Na}$), exposed to a mid-infrared laser pulse. Strong dynamical electron correlation, between the ejected electron and inner electrons, can enhance the harmonic response by orders of magnitude during the recollision. A second plateau extended far beyond the main plateau and a prominent resonance peak (for $\mathrm{Li}$) appear in the high-harmonic spectra. Although previously predicted with 1D model systems \cite{ITikhomirov,PMAbanador}, the present results establish that these effects are not an artifact of the 1D models but experimentally detectable in real systems. Our analysis quantifies the importance of electron correlation beyond the exchange interactions \cite{AGordon,SPatchkovskii,SSukiasyan} in HHG, which has been largely neglected.  Such correlations are ubiquitous in atoms, molecules, and solid-state materials. Our results have shown the high-harmonic generation provides a unique way in understanding the essential role of many-body correlations in multielectron systems exposed to external intense fields. This provides new possibilities to study inner shell electronic structure and correlations of complex molecules \cite{PMKraus2}, as well as electron-hole interaction in solid materials \cite{TIkemachi} using high-harmonic spectroscopy.

\begin{acknowledgments}
This research was supported in part by a Grant-in-Aid for Scientific Research 
(Grants No.~16H03881, No.~17K05070, and No.~18H03891)
from the Ministry of Education, Culture, Sports, Science and Technology (MEXT) of Japan.
This research was also partially supported by the Center of Innovation Program from the Japan Science 
and Technology Agency (JST), by CREST (Grant No.~JPMJCR15N1), JST, and by Quantum Leap Flagship Program of MEXT.
Y.~L. gratefully acknowledges support from JSPS Postdoctoral Fellowships for Research in Japan.
\end{acknowledgments}

\end{document}